\documentstyle[epsf]{aa}
\begin{document}
\newcommand{\AaA}{A\&A}
\newcommand{\ApJ}{ApJ}
\newcommand{\Natur}{Nature}
\newcommand{\MNRAS}{MNRAS}
\newcommand{\AaAS}{A\&As}
\newcommand{\PhRvL}{Phys. Rev. Lett.}
\def\gtsima{$\; \buildrel > \over \sim \;$}
\def\simgt{\lower.5ex\hbox{\gtsima}}
\thesaurus{}
\title{Using the COBE/DMR data as a test-bed for normality assessments}

\author{N. Aghanim\inst{1} \and O. Forni\inst{1} \and F.R. Bouchet\inst{2}}
\offprints{N. Aghanim, aghanim@ias.fr}
\institute{IAS-CNRS, Universit\'e Paris Sud, B\^atiment 121, F-91405 Orsay 
Cedex \and IAP, 98 bis Boulevard Arago, F-75014 Paris}
\date{Received date / accepted date}
\maketitle
\markboth{Using the COBE/DMR data as a test-bed for normality assessments}{}

\begin{abstract}
A very important property of a statistical distribution is to know
whether it obeys Gaussian statistics or not. On the one hand, it 
is of paramount importance in
the context of CMB anisotropy studies, since deviations from a
Gaussian distribution could indicate the presence of uncorrected
measurement systematics, of remaining fluctuations contributed by
foregrounds, of deviations from the simplest models of inflation or 
of topological defects. On the other hand, looking for a non-Gaussian
signal is a very ill-defined task and performances of various
assessment methods may differ widely when applied in different
contexts. 

In previous studies, we introduced a sensitive wavelet-based method
which we apply here to the COBE/DMR data set, already 
extensively studied, using different approaches. This provides an
objective way to compare methods. It turns out that our  multi-scale
wavelet decomposition is a sensitive method. Yet we show that the
results are rather sensitive to the choice of both the decomposition
scheme and the wavelet basis. We find that the detection of the 
non-Gaussian signature is ``marginal'' with a probability of at most 99\%. 


\end{abstract}

\section{Introduction}
The COBE-DMR data (Smoot et al. 1992) give the distribution of 
temperature anisotropies of the Cosmic Microwave Background (CMB) on
most of the sky with a resolution of about $10^\circ$. This data set was the
first to put constraints on the spectral index and the amplitude of
the initial power spectrum (Fixsen et al. 1996). Further analyses of the CMB
anisotropies should also give hints on the dominant mechanism which
generates the initial density perturbation. As a matter of fact, 
the simplest inflation models generally predict a normal (Gaussian) 
distribution of the
perturbations whereas the topological defects generate a non-Gaussian
distribution. 

Many studies aimed at characterising the statistical properties of the
COBE/DMR data 
(Amendola 1996; Kogut et al 1996; Ferreira et al. 1998; Heavens 1998; Pando
et al. 1998b; Schmalzing \& Gorski 1998; Bromley \& Tegmark 1999; Magueijo 
2000; Mukherjee et al. 2000).
These investigations have relied on various statistical indicators
both in the real space (e.g. high order moments of the temperature 
distribution, Minkowski functionals, genus analysis, ...) and in the dual,
Fourier or wavelet transformed, space (e.g. N-point correlation functions,
cumulants and moments of coefficients, ...). The COBE/DMR temperature 
fluctuations have long been considered Gaussian distributed. However more
recently, their statistical nature has been very debated, as some of the 
groups mentioned above found a non-Gaussian signature in the COBE/DMR four 
year maps. 

Regardless of the implications of such a signature, we use the
COBE/DMR data as the only available test bench for applying, on real 
maps, a method proposed in Aghanim \& Forni (1999) and Forni \& Aghanim (1999), geared at
detecting non-Gaussian features. This method is based on the
statistical analysis of the wavelet coefficients obtained after a multi-scale
decomposition of the signal. In fact, the non-Gaussian signatures are
enhanced in the wavelet space, which makes an 
analysis in this space more sensitive to the statistical signatures. 
In this context, the major differences between the methods based on wavelets
come from the choices of the decomposition scheme and the wavelet basis.
We briefly describe these choices and the method in section 2.
We present our results and discuss them in respectively section 3. and section 
4. Finally, we give our main conclusions in section 4.
%
\section{Data and analysis method}

The COBE/DMR four year data was ``cleaned'' from the contribution of
the Galactic foreground emissions using two methods. One is based on a 
combination technique (linear combination of the DMR maps) and 
results on what we refer to as the DCMB maps. The second technique is based
on the subtraction of the Galaxy contribution and it gives the DSMB
maps.

Wavelet analyses were shown to be very useful tools to detect non-Gaussian
signatures on simulated CMB maps (Aghanim \& Forni 1999; Hobson et al. 1999). They were applied
on the COBE/DMR maps by Pando et al. (1998b) and more recently in an exhaustive 
study by Mukherjee et al. (2000). In this paper, we re-analyse both sets of 
COBE/DMR data (DCMB and DSMB) to study their statistical nature (Gaussian or 
non-Gaussian). In order for our conclusions to be the most 
independent possible of the cleaning method, we have 
analysed the less contaminated regions of the COBE/DMR sky namely: the 
northern and the
southern poles.  These regions are associated respectively with the faces \#1 
and \#6 of the so-called COBE cube on which the data were projected. 

Based on the method proposed in Forni \& Aghanim (1999), we decompose the signal
on the wavelet basis and compute the kurtosis which is the fourth order 
moment of the wavelet coefficient distribution. We thus derive the excess of 
kurtosis that is the deviation from the predicted value in a Gaussian 
distribution. The values obtained
are compared with the excesses of kurtosis calculated using 2000 Gaussian 
realisations of the COBE/DMR maps. These Gaussian realisations have the same 
power spectra as the COBE/DMR maps with truly random
phases, everything else (pixel sizes, coverage) being kept the
same. Any statistical difference between the excess of kurtosis
measured on COBE/DMR and those measured on the associated Gaussian
realisations should thus be associated with the non-Gaussian character
of the data. 

The wavelet analysis consists in hierarchically decomposing an input signal 
into a series of successively lower resolution ``reference'' signals and their 
associated ``detail'' signals (Mallat 1989). At each decomposition
level, L, the 
reference signal has a 
resolution reduced by a factor of $\mathrm{2^L}$ with respect to the input 
signal. Due to the orthogonality properties of the basis, a function can be completely
characterised by the wavelet basis and the wavelet coefficients of the
decomposition. We test the detection of non-gaussianity against the 
wavelet basis in the ``dyadic'' decomposition scheme. We will compare
our results with a study based on an alternative decomposition scheme: the
``pyramidal'' decomposition (Mukherjee et al. 2000). The dyadic decomposition is a transform 
in which, at each 
level L, only the reference signal (low-pass part of the signal) is 
decomposed. At each decomposition level, the analysis is applied in both 
directions and the total number of sub-bands after L levels of decomposition 
is 3L+1. The pyramidal decomposition is similar to the previous 
decomposition in the sense that only the reference sub-band is decomposed at 
each level. However, it is performed here separately in the two directions
of the image. The total number of sub-bands after L levels of 
decomposition is then $\mathrm{(L+1)^2}$.  

The choice of the wavelet basis is critical to test for non-gaussianity. We
expect that the non-Gaussian features are due to point sources or sharp 
gradients in the signal. The detection of non-gaussianity is based on the 
search of these gradients. We therefore choose to use anti-symmetric wavelets 
which are proportional to, or almost proportional to, a first derivative 
operator and thus particularly sensitive to gradients. In practice, we choose 
the 6/10 tap filter (filter \#3) of Villasenor et al. (1995) mainly used in data 
compression. It has, in fact, a very good impulse response and a low shift 
variance which better preserve the amplitude and the location of the details.
We have also used the commonly used Haar wavelet which is also an 
anti-symmetric function. 

In the dyadic wavelet decomposition and using the anti-symmetric wavelet 
basis, we can discriminate between the 
coefficients associated with vertical, horizontal and diagonal gradients. They 
are analogous to the partial derivatives, respectively 
$\partial/\partial x$, $\partial/\partial y$ and 
$\partial^2/\partial x \partial y$ of the signal. \par

The maps we analyse have $32\times32$ pixels, with a pixel of
$2.8^\circ$ aside. Due to this reduced number, we restrict to a three
level decomposition. The three decomposition scales (hereafter I, II and III) 
correspond respectively to an angular scale $\theta=5.6^\circ$ (that
is a multipole order $\ell=35.7$), $\theta=11.2^\circ$ ($\ell=17.8$), and 
$\theta=22.4^\circ$ ($\ell=8.9$).

\section{Results}

\subsection{North galactic pole}
\begin{figure}
\epsfxsize=\columnwidth
\hbox{\epsffile{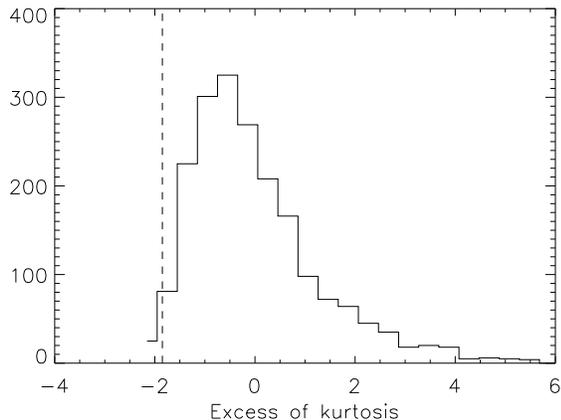}}
\caption{{\small\it For the north galactic pole of the COBE/DMR \#1 map (DCMB) 
and the Haar wavelet basis,
we show the distribution of the excesses of kurtosis computed with the
wavelet coefficients ($\partial/\partial\,y$) at the third decompostion 
scale for 2000 Gaussian realisations. The horizontal dashed line represents 
the excess of kurtosis measured on the COBE map with a probability of 98.7\%
(cf. line 8, Table \ref{tab:nph}). }}
\label{fig:disk}
\end{figure}

\begin{table}
\begin{center}
\begin{tabular}{|c|ccccc|}
\hline
Scale I & ${\overline k}$ & $\sigma_-$ & $\sigma_+$ & $k_{COBE}^1$ & P \\
\hline
$\partial/\partial x$ & 0.005 & 0.26 & 0.34 & 0.15 & 72.4\% \\
$\partial/\partial y$ & -0.006  & 0.27 & 0.36 & -0.22 & 77.6\% \\
$\partial^2/\partial x\partial y$ & 0.01 & 0.26 & 0.35 & -0.30 & 87.8\% \\
\hline
Scale II & &  & & & \\
\hline
$\partial/\partial x$ & 0.03 & 0.50 & 0.84  & -0.51 & 84.6\% \\
$\partial/\partial y$ & -0.002 & 0.48 & 0.79 &  -0.35 & 76\% \\
$\partial^2/\partial x\partial y$ & -0.02 & 0.46 & 0.76 & 1.37 & \fbox{ 97.6\%}\\
\hline
Scale III & &  & & & \\
\hline
$\partial/\partial x$ & -0.03 & 1.01 & 1.88  & -0.30 & 50.2\% \\
$\partial/\partial y$ & 0.02 & 0.99 & 1.81 &  -1.85 & \fbox{ 98.7\%} \\
$\partial^2/\partial x\partial y$ & 0.02 & 1.04 & 1.95 & 1.08 & 81.6\% \\
\hline
\end{tabular}
\end{center}
\caption{For the north galactic pole and the Haar wavelet basis, the first 
column represents the mean excess of kurtosis computed with the 
wavelet coefficients over 2000 Gaussian realisations of the COBE/DMR \#1 map 
(DCMB). $\sigma_-$ and $\sigma_+$ are the standard deviations with respect to
the mean of the kurtosis distribution. The fourth column represents
the excess of kurtosis computed on the COBE/DMR map and the last column gives
the probability, computed with the corresponding distribution, for the excess 
of kurtosis $k_{COBE}^1$ to indicate a non-Gaussian signal.}
\label{tab:nph}
\end{table}
We test for the detection of a non-Gaussian signature in the COBE/DMR, DCMB 
map, of the north galactic pole against 2000 Gaussian realisations of this 
map. First using the Haar wavelet basis in a three level decomposition, we 
compute the excesses of kurtosis
of the coefficients associated with the details for the COBE map 
($k_{COBE}^1$, see Table~\ref{tab:nph}), and the Gaussian
realisations. We give, in this case, the mean  
excess of kurtosis ${\overline k}$ and the standard deviations $\sigma_-$ and 
$\sigma_+$ with respect to this mean for the quantities that are respectively
smaller and larger than ${\overline k}$. Finally, we compare the obtained 
results and quantify the probability P that the measured excess of kurtosis on 
the COBE map ($k_{COBE}^1$) indicates a non-Gaussian process. This 
probability is computed by integrating the excess of kurtosis distribution
for the Gaussian realisations, one example of which is shown in Figure 
\ref{fig:disk}. It ranges between about 50 and 99\% depending on the 
decomposition scale and on the type of details we study. We note that the
non-Gaussian signature is detected at all scales with at least a probability
larger than 68\%. However, the most significant detections at about
the 98 and 99\% confidence levels are obtained respectively at the second and
third scales where the beam and noise effects are expected to be less
dominant.  In the case of the Villasenor et al. wavelet basis, the  
probability of detecting a non-Gaussian signature in the DCMB COBE/DMR \#1 map 
lies between about 67 and 98\% (see Table~\ref{tab:npv}). Again this detection is made at all scales.\par

We perform the same analysis using the two wavelet basis on the COBE map 
obtained with the second subtraction technique (DSMB maps). In Table 
\ref{tab:dsm}, we give the probability for the excess of kurtosis, measured on 
the COBE/DMR map, to be
associated with the detection of a non-Gaussian signal. We note again that the
probabilities indicate a non-Gaussian process at all scales. The Haar
basis seems however to give smaller (less discriminating) values than the Villasenor et al. basis.
\begin{table}
\begin{center}
\begin{tabular}{|c|ccccc|}
\hline
Scale I & ${\overline k}$ & $\sigma_-$ & $\sigma_+$ & $k_{COBE}^1$ & P \\
\hline
$\partial/\partial x$ & 0.004 & 0.26 & 0.30 & 0.75 & \fbox{ 98\%} \\
$\partial/\partial y$ & 0.01 & 0.27 & 0.39 & 0.75 & \fbox{ 97.7\%} \\
$\partial^2/\partial x\partial y$ & 0.007 & 0.27 & 0.36 & -0.18 & 73.3\% \\
\hline
Scale II & &  & & & \\
\hline
$\partial/\partial x$ & 0.01 & 0.48 & 0.78  & 0.59 & 87.5\% \\
$\partial/\partial y$ & 0.03 & 0.50 & 0.87 & -0.41 & 79.3\% \\
$\partial^2/\partial x\partial y$ & 0.04 & 0.50 & 0.83 & 0.79 & 89.9\% \\
\hline
Scale III & &  & & & \\
\hline
$\partial/\partial x$ & -0.01 & 1.0  & 1.79 & -0.77 & 66.8\% \\
$\partial/\partial y$ & 0.02 & 1.02 & 1.83 & -1.17 & 86.3\% \\
$\partial^2/\partial x\partial y$ & 0.005 & 0.99 & 1.83 & -1.59 & \fbox{ 95.6\%} \\
\hline
\end{tabular}
\end{center}
\caption{Same as Table~\ref{tab:nph}, for the the Villasenor et al. wavelet 
basis.
\label{tab:npv}}
\end{table}

\subsection{South galactic pole}

Using the power spectrum of the COBE/DMR \#6 map (DCMB) of the south galactic 
pole, we again simulate 2000 Gaussian realisations of this map. We
perform the wavelet decomposition using the Haar and the
Villasenor et al. bases, and analyse the statistical properties of
each type of details. As for the north pole, we
compute the excess of kurtosis in the COBE/DMR map ($k_{COBE}^6$) and the 
probability, P, that it belongs to the distribution of the excesses of
kurtosis of the Gaussian realisations. The latter is characterised by its mean 
${\overline k}$ and standard deviations $\sigma_-$ and $\sigma_+$ of 
respectively smaller and larger values with respect to the mean. The
results are summarized in Tables~\ref{tab:sph} and \ref{tab:spv}.
\begin{table}
\begin{center}
\begin{tabular}{|c|ccccc|}
\hline
Scale I & ${\overline k}$ & $\sigma_-$ & $\sigma_+$ & $k_{COBE}^6$ & P \\
\hline
$\partial/\partial x$ & 0.006 & 0.26 & 0.35 & -0.28 & 85.7\%\\
$\partial/\partial y$ & -0.01 & 0.26 & 0.36 & -0.21 & 75.8\% \\
$\partial^2/\partial x\partial y$ & 0.02 & 0.27 & 0.35 & -0.28 & 85.7\% \\
\hline
Scale II & &  & & & \\
\hline
$\partial/\partial x$ & 0.03 & 0.50 & 0.82 & -0.56 & 87.6\% \\
$\partial/\partial y$ & 0.003 & 0.49 & 0.80 & 0.59 & 87.7\% \\
$\partial^2/\partial x\partial y$ & -0.03 & 0.46 & 0.75 & -0.62 & 91.5\% \\
\hline
Scale III & &  & & & \\
\hline
$\partial/\partial x$ &-0.03 & 1.01 & 1.89 & -0.68 & 67.5\% \\
$\partial/\partial y$ & 0.007 & 0.99 & 1.81 & -1.27 & 90.3\% \\
$\partial^2/\partial x\partial y$ & 0.005 & 1.03 & 1.93 & 0.40 & 52.9\% \\
\hline
\end{tabular}
\end{center}
\caption{For the south galactic pole and the Haar wavelet basis, the first
column represents the mean excess of kurtosis computed with the 
wavelet 
coefficients for 2000 Gaussian realisations of the COBE/DMR \#6 map (DCMB).
$\sigma_-$ and $\sigma_+$ are the standard deviations with respect to the mean 
of the kurtosis distribution. The fourth column represents
the excess of kurtosis computed on the COBE/DMR map and the last column gives
the probability, computed with the corresponding distribution, for the excess 
of kurtosis $k_{COBE}^6$ to be associated with a non-Gaussian signal.}
\label{tab:sph}
\end{table}
\begin{table}
\begin{center}
\begin{tabular}{|c|ccccc|}
\hline
Scale I & ${\overline k}$ &  $\sigma_-$ & $\sigma_+$ & $k_{COBE}^6$ & P \\
\hline
$\partial/\partial x$ & 0.006 & 0.27 & 0.35 & -0.003 & 48.6\%\\
$\partial/\partial y$ & 0.01 & 0.26 & 0.38 & 0.33 & 86.2\% \\
$\partial^2/\partial x\partial y$ & 0.006 & 0.27 & 0.36 & 0.11 & 68.7\% \\
\hline
Scale II & &  & & & \\
\hline
$\partial/\partial x$ & 0.01 & 0.48 & 0.78 & -0.55 & 86.4\% \\
$\partial/\partial y$ & 0.03 & 0.50 & 0.90 & 0.10 & 62.8\% \\
$\partial^2/\partial x\partial y$ & 0.04 & 0.50 & 0.82 & 0.61 & 86.9\% \\
\hline
Scale III & &  & & & \\
\hline
$\partial/\partial x$ &-0.01 & 0.99 & 1.81 & -0.21 & 46.9\% \\
$\partial/\partial y$ & 0.04 & 1.02 & 1.89 & -1.26 & 87.5\% \\
$\partial^2/\partial x\partial y$ & 0.002 & 1.0 & 1.81 & 0.45 & 73.3\% \\
\hline
\end{tabular}
\end{center}
\caption{Same as Table \ref{tab:sph} for the Villasenor et al. wavelet 
basis.}
\label{tab:spv}
\end{table}

\begin{table*}
\begin{center} 
\begin{tabular}{|c|c|c|c|c||c|c|c|} 
\hline
Wavelet basis & Scale & \multicolumn{3}{|c||}{North pole} & \multicolumn{3}{|c|}{South pole} \\ 
\hline
 & & \multicolumn{1}{|c|}{$\partial/\partial x$} & \multicolumn{1}{|c|}{$\partial/\partial y$} & \multicolumn{1}{|c|}{$\partial^2/\partial x\partial y$}& \multicolumn{1}{|c||}{$\partial/\partial x$} & \multicolumn{1}{|c|}{$\partial/\partial y$} & \multicolumn{1}{|c|}{$\partial^2/\partial x\partial y$}   \\
\hline 
Haar & I & 66.3\% & \fbox{ 99\%} & 58.9\%  & 64.1\% & 81.5\% & 86.5\%  \\ 
     & II  & \fbox{ 97.5\%} & 83.9\% & 76.2\%  & 83.2\%& 85.4\% & 52.4\%  \\
     & III & 57.5\% & 71.9\% & 79.9\% & 79.5\% & 73.4\% & 43.6\%  \\
\hline
\hline
Villasenor et al.& I & \fbox{ 99.3\%} & \fbox{ 98.1\%} & 92.2\% & 46.9\% & 86\% & 63\%  \\
           & II & 81.8\% & 50.8\% & 77.7\% & \fbox{ 98.4\%} & 92\% & 65.4\%  \\
	   & III & 68\% & 82\% & 94.9\% & 55.1\% & 55.9\% & 88.9\%  \\
\hline
\end{tabular}\\
\end{center}
\caption{Probability for the excess of kurtosis of the COBE/DMR map (DSMB) to
be attributed to a non-Gaussian signature. The numbers are given, for the three types of
details, as a function of the decomposition scale and the wavelet basis. The
first part of the table represents the north pole and the second part is for
the south pole.}
\label{tab:dsm}
\end{table*}
At first sight, these results suggest that the southern hemisphere
distribution is more Gaussian than its northern hemisphere
counterpart. As a matter of fact, the probability that the excess of
kurtosis $k_{COBE}^6$ measured on the COBE DCMB map indicates a 
non-Gaussian process lies between 53 and 92\% for the Haar wavelet
basis and  between 47 and 87\% for the Villasenor et al. wavelet. At
all scales, we measure a probability larger than 68\% for at least one
type of details. However, these values are always smaller than those
obtained for the northern hemisphere.

When we analyse the DSMB map of the southern hemisphere (Table~\ref{tab:dsm}),
we note the same behaviour: the non-Gaussian signature is weaker in
the south than in the north. The probabilities lie between 44 and 87\%
for the Haar wavelet basis, and between 45 and 98\% for the Villasenor
et al. basis. The highest probabilities, in this case, are obtained at
the second decomposition scale.
\subsection{Scale-scale correlation spectra}
\begin{table*}
\begin{center} 
\begin{tabular}{|c|c|c|c|c||c|c|c|} 
\hline
Wavelet basis & Scale-scale & \multicolumn{3}{|c|}{North pole} & \multicolumn{3}{|c|}{South pole} \\ 
\hline
 & & \multicolumn{1}{|c|}{$\partial/\partial x$} & \multicolumn{1}{|c|}{$\partial/\partial y$} & \multicolumn{1}{|c|}{$\partial^2/\partial x\partial y$} & \multicolumn{1}{|c|}{$\partial/\partial x$} & \multicolumn{1}{|c|}{$\partial/\partial y$}  & \multicolumn{1}{|c|}{$\partial^2/\partial x\partial y$} \\
\hline 
Haar & I--II & 65\% & 75.5\% & 70.9\% & 79.8\% & 60\% & 86.1\%  \\ 
     & II--III & 68.7\% & 86.9\% & \fbox{ 96.1\%} & 77.6\% & 44\% & 67.1\% \\
Villasenor et al. & I--II & \fbox{ 99.3\%} & 70.1\% & 86\% & 63.4\% & \fbox{ 95.2\%} & 80\% \\
           & II--III & 61.4\% & 88.3\% & 80.7\% & \fbox{ 96.5\%} & 83\% & 49.5\% \\
\hline
\hline
Haar & I--II & 74.5\% & 73.4\% & 63.9\% & 93.5\% & 82.2\% & 67.2\%  \\ 
     & II--III & 83.8\% & 93.9\% & \fbox{ 96.4\%} & 86.6\% & 72.3\% & 68.1\% \\
Villasenor et al. & I--II & 62.4\% & 80.6\% & 79\% & 76.7\% & 71.9\% & \fbox{ 98\%} \\
           & II--III & 73.5\% & 80.2\% & 81.2\% & 83.3\% & 71.5\% & 57.3\% \\
\hline
\end{tabular}\\
\end{center}
\caption{Scale-scale correlation coefficients for the three types of details
as a function of the wavelet basis. For the south and north pole, we give the
probability for the correlation coefficient to be attributed to a non-Gaussian 
signal. The correlations are performed between the first and second scales, 
and between the second and
third scales. The upper part of the table is given for 
the DSMB COBE/DMR maps whereas the lower part is for the DCMB maps. }
\label{tab:corel}
\end{table*}
Another method to detect a non-Gaussian signature consists in using the 
scale-scale 
correlations as proposed by \cite{pando98a}, \cite{pando98} and 
\cite{mukherjee2000}. The
correlations between structures at adjacent decomposition
scales are quantified. They can be computed using the expression given in 
\cite{mukherjee2000}:
\[ C_{j_1,j_2}=\]
\begin{equation}
\frac{2^{k_1+k_2}\sum^{2^{k_1}-1}_{l_1=0}\sum^{2^{k_2}-1}_{l_2=0}<b^2_{j_1,j_2,\frac{l_1}{2},\frac{l_2}{2}}b^2_{k_1,k_2,l_1,l_2}>}{\sum_{l_1=0}^{2^{k_1}-1}\sum_{l_2=0}^{2^{k_2}-1}b^2_{j_1,j_2,\frac{l_1}{2},\frac{l_2}{2}}\sum_{l_1=0}^{2^{k_1}-1}\sum_{l_2=0}^{2^{k_2}-1}b^2_{k_1,k_2,l_1,l_2}}.
\label{eq:cor}
\end{equation}
In this expression, the $b_{j_1,j_2,l_1,l_2}$ represent the
coefficients in the  wavelet decomposition. $C_{j_1,j_2}$ gives the
correlation factor between the coefficients in the domain ($j_1,j_2$)
and its adjacent scale 
($k_1,k_2$), where $k_1=j_1+1$ and $k_2=j_2+1$, 
for a Gaussian signal $C_{j_1,j_2}=1$. 
In order to assess the gain brought by this further step of analysis,
we proceed as previously. We first compute the
scale-scale correlation coefficients for our 2000 Gaussian realisations
of the COBE/DMR maps after decomposing them on three scales with the
Haar and Villasenor et al. bases. 
We then compute the scale-scale correlations for the ``real'' COBE maps and
finally compute the probability that the measured numbers indicate a
non-Gaussian signature by using the distribution of correlations for
the Gaussian maps.

The results are summarized in Table \ref{tab:corel}. They basically
agree with those obtained by \cite{mukherjee2000} when the comparison can be
made, that is for the diagonal details. The probabilities for the correlations
between scales I and II in the north galactic pole lie between 64 and 76\%,
with the Haar basis, and between 62 and 99\% with the Villasenor et al. basis.
When we correlate scales II and III, we find a probability of 69 to 96\%, with
Haar; whereas it is of 61 to 88\% with the Villasenor et al. basis. In the
south galactic pole, the probability for the correlation coefficients between
scales I and II to be associated with a non-Gaussian signal ranges between 60
and 94\% with the Haar basis, and between 63 and 98\% with the Villasenor et
al. basis. Correlating scale II and scale III, the probabilities are between
44 and 87\% with Haar and between 50 and 97\% with Villasenor et al. The 
results obtained with the correlation coefficients agree with the direct
analysis of the wavelet coefficients. However, they seem to give larger
non-Gaussian signatures for the south pole than the direct analysis.
%
\section{Discussion and conclusions}
At each decomposition scale, we detected a non-Gaussian signature using 
three different criteria related to the horizontal, vertical and 
diagonal details of the image. We tested and compared the sensitivity to 
non-Gaussian features of the south and north pole maps obtained 
after subtracting the Galaxy with two techniques. We thus analysed
this four map set with two wavelet bases used in a three level dyadic
decomposition. All the results are summarized in Tables \ref{tab:nph}, 
\ref{tab:npv}, \ref{tab:sph}, \ref{tab:spv}, \ref{tab:dsm}. 

This rather exhaustive study leads to two main conclusions. One is
that the analysis of the COBE maps is rather sensitive to the choice
of the wavelet basis when the same decomposition scheme is used. The
second conclusion is that the non-Gaussian signature is detected with
a probability of 99\%. At the first and second decomposition scales,
the high probalities are very likely associated with
some instrumental effects, such as pixel to pixel correlations or Poisson 
noise, or systematic effects (Banday et al. 2000). At the third decomposition 
scale (which is less affected by the instrumental effects), we detect a 
non-Gaussian signature with a probability $<$ 95\% in the southern hemisphere,
whereas in the northern hemisphere, it is detected with a probability 
close to 99\%.
We have compared our results to those obtained by \cite{mukherjee2000}. 
Namely,
we compare the results they obtain for what they note $k=32$, 16 and 8, which
correspond exactly to what we refer to as the diagonal details for 
respectively the first, second and third
decomposition scales. We used the dyadic decomposition
advocated by \cite{aghanim98,forni99,aghanim99} rather than the
pyramidal decomposition used in \cite{hobson99} and in \cite{mukherjee2000}. 
This choice is motivated by the fact that this particular
decomposition scheme  gives the correlations between the two
directions of the map at each scale. It is not the case with the
pyramidal decomposition which treats both directions  
as if they were independent. Another advantage of the dyadic decomposition is
that, for each decomposition scale, the analysis bears upon the maximum
number of coefficients possible. This is crucial for statistics in
general and for the small COBE maps (\#1 and 6) in particular. 

The comparison between our results, obtained with the Villasenor et al. 
wavelet basis, and those of \cite{mukherjee2000} shows that our non-gaussianity
detection technique gives every time higher confidence levels
 except for two cases which 
are the first decomposition scale of the southern hemisphere (DCMB and DSMB 
maps). When we use the Haar wavelet basis, our confidence levels are 
higher than, or 
similar to, those quoted in \cite{mukherjee2000} except for two cases that are the
third decomposition scale of the northern and southern hemisphere of the DSMB
map. Of course in addition to the diagonal details, we have two other criteria 
to detect the 
non-Gaussian signature (vertical and horizontal details). In particular, for 
the cases where the probability is smaller than in \cite{mukherjee2000} we 
find a non-Gaussian signature
based on the vertical or/and the horizontal details.

In our study, we find that the Villasenor et al. wavelet basis combined
with a dyadic decomposition gives higher probabilities than the pyramidal 
decomposition. This basis is also more sensitive than the Haar
basis. The better performances in detecting gradients come from the larger size of its kernel (Villasenor et al. 1995). However, this 
could become a weakness when a small number of pixels is available (small maps 
or high decomposition levels). In this case, the Haar wavelet basis could be 
more adapted due to the smaller size of its kernel. 

Our choice of wavelet basis and decomposition scheme confirms that a
non-Gaussian 
signature can be detected either in the south or the north galactic pole. It
seems however that the statistical signature is less significant in the
southern pole. In this case, the largest probability, $>95$\%, is obtained only at the
second decomposition scale for the DSMB COBE/DMR \#6 map. The non-Gaussian 
signature is more significant for the north galactic pole as the probability
is often $\geq 95$\% and can even reach 99\%. 
The scale-scale correlations represent a step
further in the statistical analysis of the signal. However in our case, it does
not make a significant improvement in the results. The non-Gaussian signature
is indeed detected with a probability of 99\% at most, such as in the direct
analysis.

The large size of the beam and pixels (respectively 7 and 2.8$^\circ$) results
in a rather small number of wavelet coefficients. This, together with the low 
signal to noise ratio of the COBE/DMR data, represent very limiting factors 
which prevent us from drawing any hasty conclusion from the rather
marginal detections of the non-Gaussian signatures. This kind of statistical 
analyses will soon demonstrate all their performances when applied
to the new generation of CMB data sets of higher sensitivity and resolution
(Boomerang (De Bernardis et al. 2000) and Maxima (Hanany et al. 2000)). We thus 
expect that these studies will then
put constraints on the primordial non-Gaussian contributions such as the 
topological defects suggested for example in \cite{bouchet2000}.

\begin{acknowledgements}
This work is dedicated to the memory of our colleague Richard Gispert.
We wish to thank G. Lagache for her help with the COBE data and E. 
Martinez-Gonzales for fruitful discussions. We also thank an anonymous referee
for his helpful comments and suggestions.
\end{acknowledgements}

\end{document}